\documentclass[12pt]{article}
\pdfoutput=1
\usepackage{hyperref}
\usepackage{graphicx}
\usepackage{graphics}
\usepackage{dsfont}
\usepackage{epsfig}
\usepackage{amsmath,amssymb,amsthm,amscd}
\newcommand{\Tr}{\textrm{Tr}}

\def\ket{\rangle}
\def\bra{\langle}

\numberwithin{equation}{section}

\begin{document}
\begin{titlepage}
{}~ \hfill\vbox{ \hbox{} }\break

\rightline{USTC-ICTS-19-24}

\vskip 3 cm

\centerline{\Large 
\bf  
Note on $S$-channel factorization in multitrace} 
\vskip 0.2 cm
\centerline{\Large 
\bf 
 Berenstein-Maldacena-Nastase correlators}

\vskip 0.5 cm

\renewcommand{\thefootnote}{\fnsymbol{footnote}}
\vskip 30pt \centerline{ {\large \rm Min-xin Huang
\footnote{minxin@ustc.edu.cn}  } } \vskip .5cm  \vskip 20pt 

\begin{center}
{Interdisciplinary Center for Theoretical Study,  \\ \vskip 0.1cm  University of Science and Technology of China,  Hefei, Anhui 230026, China} 
 \\ \vskip 0.3 cm
{NSFC-SFTP Peng Huanwu Center for Fundamental Theory,  \\ \vskip 0.1cm  Hefei, Anhui 230026, China} 
\end{center}

\setcounter{footnote}{0}
\renewcommand{\thefootnote}{\arabic{footnote}}
\vskip 60pt
\begin{abstract}

In a previous paper we proposed a factorization principle for the correlation functions of Berenstein-Maldacena-Nastase (BMN) operators in free $\mathcal{N}=4$ super-Yang-Mills theory. These correlators are conjectured to described physical string amplitudes in an infinitely curved Ramond-Ramond pp-wave background. There was a puzzle that the factorization seems to break down for $S$-channel in the $2\rightarrow 2$ scattering process. Here we resolve this puzzle by including some diagrams missed in the previous paper. We also observe some interesting relations which further support the interpretation of higher genus correlators as physical string loop amplitudes.

\end{abstract}

\end{titlepage}
\vfill \eject

%%%%%%%%%%%%%%%%%%%%%%%%%%%%%%%%%%%%%%%%%%%%%%%%%%%%%%%%%%%%%

\newpage

\baselineskip=16pt

\tableofcontents

\section{Introduction and Summary}

In string theory realizations of particle physics models, we usually assume the string length scale is very small, or the string energy scale is very high. The stringy excitation modes are not observable in low energy, and the ground states of string mode give rise to the various fundamental particles that we know. 

However,  if string theory is consistent at arbitrary energy scale, it is interesting to also consider the opposite limit where the string length is very long. The motivation is probably not for the purpose of constructing realistic particle physics model, but for purely theoretical reasons, for example a deeper understanding of the still mysterious AdS/CFT correspondence \cite{Maldacena, Gubser, Witten}, which states that type IIB string theory on the $AdS_5\times S^5$ background is equivalent to the maximally supersymmetric $SU(N)$ gauge theory in four dimensions.  A particularly interesting limit is the pp-wave geometry, a Penrose limit zooming in the null geometry of the  $AdS_5\times S^5$ spacetime. The string excitation modes are described by the Berenstein-Maldacena-Nastase (BMN) operators on the field theory side \cite{BMN}.   

In our previous papers \cite{Huang:2002, Huang:2010} we studied the (higher genus) correlation functions of BMN operators in free gauge theory. The field theory side of the correspondence is easy. However, on the string theory side, this corresponds to an infinitely negatively curved Ramond-Ramond pp-wave background, where strings are effectively infinitely long and tensionless, and all stringy excitation modes have degenerate mass. Usually the effective field theory approach breaks down in this scenario, and we cannot say much about the underlying physics. However, it seems somehow luckily the stringy physics also becomes extremely simplified. We proposed  that the string amplitudes can be computed simply by cubic diagrams, and there is a so-called ``factorization" principle relating the string diagram calculations and field theory calculations, in the spirit of AdS/CFT correspondence. Here we do not have a Lagrangian description on the string side because of the infinite string length. The nice situation arises, probably due to the fact that the spacetime is highly compressed by the infinite curvature. Its structure becomes effectively that of a single point and is thus extremely simple. The string diagrams have only cubic vertices, but no propagator between the vertices, signaling string interactions occurring instantaneously without mediation in an ambient spacetime.  The general form of the factorization rule is 
\begin{eqnarray}
S_i = \sum_j m_{ij} F_j,
\end{eqnarray}
where $S_i$ and $F_j$ denote string and field theory diagram contributions, and $m_{ij}$ are non-negative integers denoting the multiplicity of expanding the ``short process" of field theory diagrams into the ``long process" of string diagrams. 

If our claim is valid, the higher genus correlation functions represent string loop amplitudes in this simple background. We can straightforwardly compute higher string loop amplitudes which are notoriously difficult. In flat space for critical string theories most calculations have been restricted to less than two loops; see, e.g.,   \cite{DHoker:1988pdl}. We can compare with other special situations where higher genus string amplitudes are computable. One case is the noncritical string theories dual to matrix models or matrix quantum mechanics, intensely studied in early 1990s; see, e.g., \cite{Ginsparg:1993is}. Another case is topological string theory, where all genus partition functions are computed for a certain class of noncompact Calabi-Yau manifolds, by, e.g., the topological vertex method \cite{Aganagic:2003db}. For compact Calabi-Yau manifolds, one can compute the amplitudes to very high but not all genera, e.g., in \cite{Huang:2006hq, Huang:2015sta}.  In these special situations the physical string degrees of freedom are rather simple, mostly ``minimal" or topological in nature, lacking the infinite stringy oscillator modes. Here although the spacetime structure of the infinitely curved pp-wave background  is extremely simple, we still have the rich physical degrees of freedom of infinite stringy oscillator modes of conventional critical string theory. In this case there is no apparent technical obstruction to compute string amplitudes to any higher genus. 

We think it is worthwhile to revisit the proposal in our previous papers \cite{Huang:2002, Huang:2010} from time to time with fresh perspective, and to make incremental improvements and corrections. Many interesting developments over the years seem to support the physical significance of infinite curvature limit in AdS/CFT correspondence, and are potentially relevant for further explorations of our work. Gopakumar has long promoted the studies of free field theory in the context of AdS/CFT correspondence in a series of papers \cite{Gopakumar:2003ns}. Many useful techniques have been developed to compute correlation function of large charge operators; see, e.g., \cite{Pasukonis:2010rv, Berenstein:2019esh}. More recently, some progress has been  made to understand string perturbation theory in pp-wave background in the honest Neveu-Schwarz-Ramond formalism \cite{Cho:2018nfn}.  Also, Berkovits uses pure spinor formalism to understand free super-Yang-Mills Feynman diagrams  in the the small AdS radius limit \cite{Berkovits:2019ulm}.

Since this is a short note, we should not provide a lengthy review of the topic here. Basic properties of BMN operators are studied in the early pioneering papers \cite{Constable1, KPSS}. We use the notations in our previous paper \cite{Huang:2010}.  In the current work, in Sec. \ref{sec2}, we clarify a puzzle about $S$-channel factorization in our previous paper.   in Sec. \ref{sec3}, we comment on some properties of the  higher genus BMN correlators.

\section{The $2\rightarrow 2$ scattering process}
\label{sec2}
The $2\rightarrow 2$ scattering process is particularly familiar from collider physics. Since our spacetime is collapsed to a point, the process here should be thought of as instantaneous, instead of scattering to and from space infinity as in usual quantum field theory in flat space. To be consistent we use the same symbols for various diagrams in our previous paper \cite{Huang:2010}. We also consider three cases, and point out the missing diagrams in each case. For convenience we list the formulas for BMN operators 
 \begin{eqnarray} \label{BMNoperators}
&& O^{J}  = \frac{1}{\sqrt{JN^J}}TrZ^J,  ~~~~~~~~~
O^{J}_{0} = \frac{1}{\sqrt{N^{J+1}}} Tr(\phi^{I} Z^{J}), \nonumber \\
&& O^J_{-m,m} = \frac1{\sqrt{JN^{J+2}}} \sum_{l=0}^{J}e^{\frac{2\pi iml}{J}}
Tr(\phi^{I_1} Z^l\phi^{I_2} Z^{J-l}).
\end{eqnarray}
Here $Z$ is a complex scalar in the $\mathcal{N}=4$ super-Yang-Mills theory, and $\phi^{I_1}$ and $\phi^{I_2}$ are the two different real scalar fields out of the four remaining ones besides those in $Z$.  We take the BMN limit $J, N \rightarrow \infty$ with  finite $g:=\frac{J^2}{N}$ as the effective string coupling constant. The planar three point functions can be derived from the Green-Schwarz string field theory vertex \cite{SV, Huang1}, and serve as the cubic vertex in the string diagrams. As in the previous paper, we omit the universal spacetime factor in the correlators. 

\begin {eqnarray} \label{planar1}
&& \langle\bar{O}^JO^{J_1}O^{J_2}\rangle=\frac{g}{\sqrt{J}}\sqrt{x(1-x)}, ~~~
\langle\bar{O}^J_{0}O^{J_1}O^{J_2}_{0}\rangle = \frac{g}{\sqrt{J}}x^{\frac{1}{2}}(1-x), \nonumber \\
&& \langle\bar{O}^J_{00}O^{J_1}_{0}O^{J_2}_{0}\rangle = \frac{g}{\sqrt{J}}x(1-x), ~~~
\langle\bar{O}^J_{00}O^{J_1}_{00}O^{J_2}\rangle = \frac{g}{\sqrt{J}}x^{\frac{3}{2}}(1-x)^{\frac{1}{2}},  \nonumber \\
&& \langle\bar{O}^J_{-m,m
}O^{J_1}_{0}O^{J_2}_{0}\rangle = - \frac{g}{\sqrt{J}}\frac{\sin^2(\pi mx)}{\pi^2m^2},  \nonumber \\
&& \langle\bar{O}^J_{-m,m}O^{J_1}_{-n,n}O^{J_2}\rangle = \frac{g}{\sqrt{J}}x^{\frac{3}{2}}(1-x)^{\frac{1}{2}}\frac{\sin^2(\pi mx)}{\pi^2 (mx-n)^2},
\end {eqnarray}
where $x=\frac{J_1}{J}$,  and it is implicit that $J=J_1+J_2$. If the denominator is $0$ in the last two formulas, then one uses special case formulas. The following integral form is valid for all integers $m,n$, and is sometimes quite useful for checking the factorization rule  
\begin{eqnarray} \label{integralform}
\langle\bar{O}^J_{-m,m
}O^{J_1}_{0}O^{J_2}_{0}\rangle&=& \frac{g}{\sqrt{J}} (\int_0^x dy_1 e^{2\pi i m y_1})  (\int_x^1 dy_2 e^{-2\pi i m y_2}) \\ \nonumber 
\langle\bar{O}^J_{-m,m
}O^{J_1}_{-n,n}O^{J_2}\rangle&=& \frac{g}{\sqrt{J}} ( \frac{1-x}{x} )^{\frac{1}{2}}  (\int_0^x dy_1 e^{2\pi i (m-\frac{n}{x})y_1})  (\int_0^x dy_2 e^{-2\pi i (m-\frac{n}{x})y_2}). 
\end{eqnarray}

\subsubsection{Case one: $\bra \bar{O}^{J_1} \bar{O}^{J_4}O^{J_2}O^{J_3}\ket$}

\begin{figure}
  \begin{center}
  \includegraphics[width=6.5in]{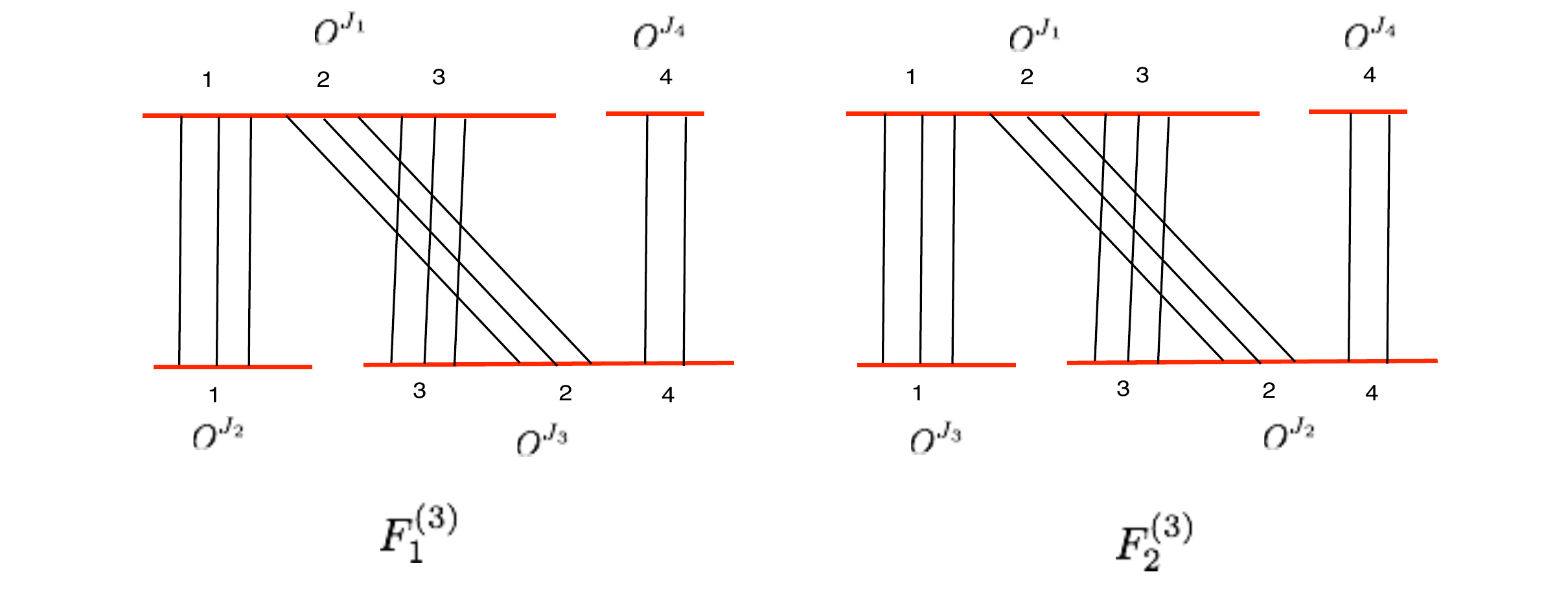} 
 \end{center}
\caption{Some field theory diagrams for  $\bra \bar{O}^{J_1} \bar{O}^{J_4}O^{J_2}O^{J_3}\ket$.}  \label{F3}
\end{figure}

This is the correlator of the vacuum operators, and it is implicit that $J_1+J_4=J_2+J_3$, and we denote $x_i=\frac{J_i}{J}$. Without loss of generality we assume $J_1>J_2>J_3>J_4$.  In the previous paper we calculated the field theory amplitudes for diagrams depicted in Fig. \ref{F3} and found
\begin{eqnarray}  
F^{(3)}_1 &=& \frac{J_1J_2J_3J_4(J_1-J_2)}{N^2\sqrt{J_1J_2J_3J_4}}= \frac{g^2}{J}(x_1x_2x_3x_4)^{\frac{1}{2}}(x_1-x_2),  \nonumber \\
F^{(3)}_2 &=& \frac{J_1J_2J_3J_4(J_1-J_3)}{N^2\sqrt{J_1J_2J_3J_4}}= \frac{g^2}{J}(x_1x_2x_3x_4)^{\frac{1}{2}}(x_1-x_3). \label{F3eq} 
\end{eqnarray}

\begin{figure}
  \begin{center}
  \includegraphics[width=3.3in]{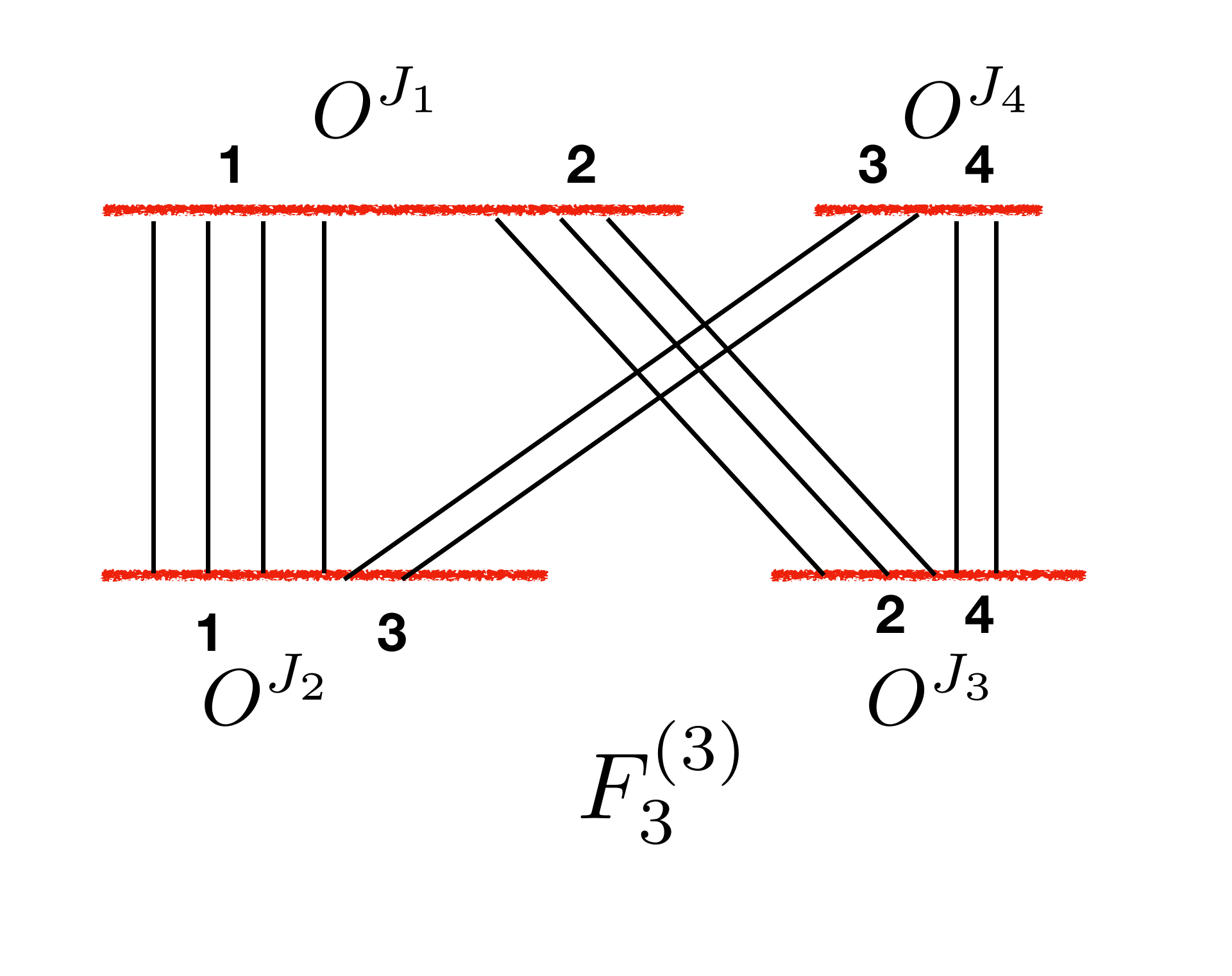} 
 \end{center}
\caption{The new diagram missed in our previous paper for  $\bra \bar{O}^{J_1} \bar{O}^{J_4}O^{J_2}O^{J_3}\ket$.}  \label{F3a}
\end{figure}

The missing diagram in the previous paper \cite{Huang:2010} is depicted in Fig. \ref{F3a}. Here the double trace is also further dissected into four segments, and one can check it has the same large $N$ factor as the diagrams in Fig. \ref{F3}. To compute its contribution, we note that the diagram is determined by dissecting the shortest trace $O^{J_4}$ into two parts, and then the positions of the other dissecting points uniquely follow. This contributes a factor of $J_4$ and the other factors are the same as in the (\ref{F3eq}). So the contribution is 
\begin{eqnarray}  
F^{(3)}_3 &=& \frac{J_1J_2J_3J_4}{N^2\sqrt{J_1J_2J_3J_4}} J_4= \frac{g^2}{J}(x_1x_2x_3x_4)^{\frac{1}{2}}x_4.
\end{eqnarray}

\begin{figure}
  \begin{center}
  \includegraphics[width=6.5in]{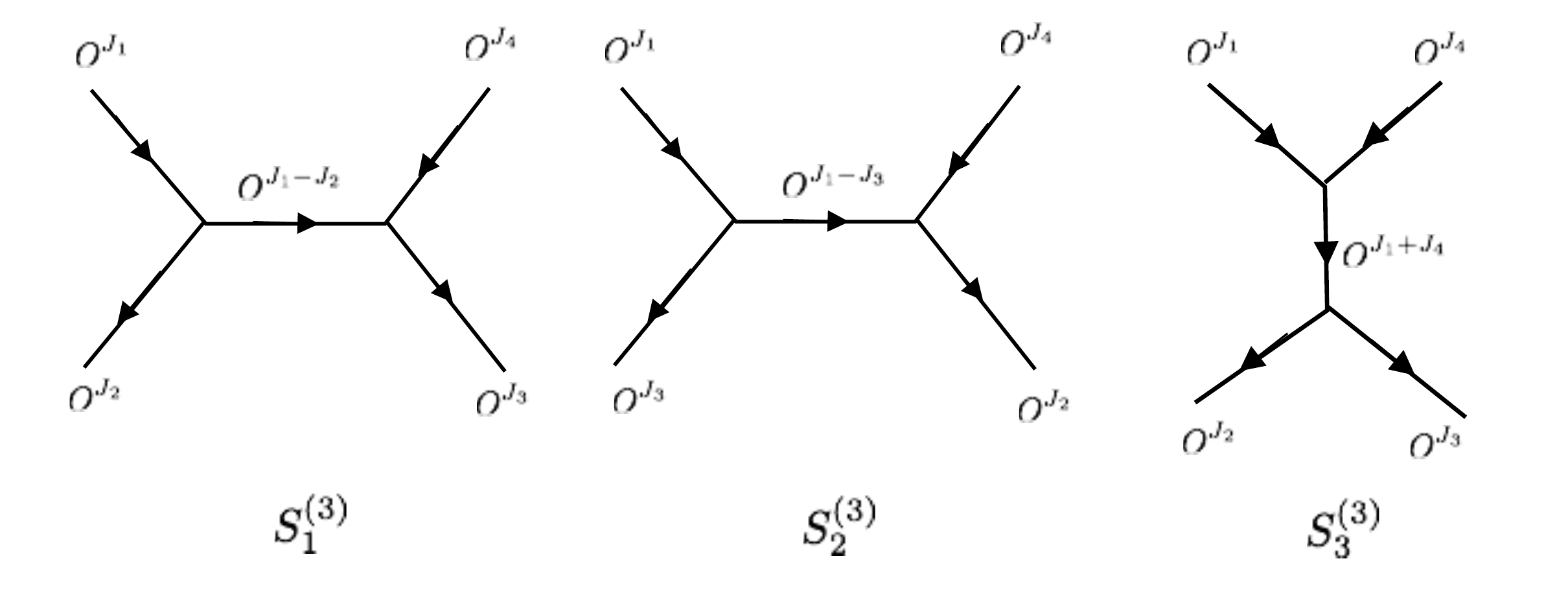} 
\end{center}
\caption{The string diagrams for  $\bra \bar{O}^{J_1} \bar{O}^{J_4}O^{J_2}O^{J_3}\ket$. We denote the contributions of the three diagrams $S^{(3)}_1$,  $S^{(3)}_2$, $S^{(3)}_3$ respectively. The three diagrams represent the $T$, $U$, $S$ channels in $2\rightarrow 2$ scattering. }  \label{S3}
\end{figure}

The string diagrams are depicted in Fig. \ref{S3}, and their contributions are 
\begin{eqnarray}
S^{(3)}_1 &=&  \frac{g^2}{J}(x_1x_2x_3x_4)^{\frac{1}{2}}(x_1-x_2) \\
S^{(3)}_2 &=&  \frac{g^2}{J}(x_1x_2x_3x_4)^{\frac{1}{2}}(x_1-x_3) \\
S^{(3)}_3 &=&  \frac{g^2}{J}(x_1x_2x_3x_4)^{\frac{1}{2}}
\end{eqnarray}

We follow the proposal in our previous paper \cite{Huang:2010} to count the multiplicity, by expanding the short process of the field theory diagrams into long processes 
\begin{eqnarray}
F^{(3)}_1 : && (123)_1(4)_4 \rightarrow (1)_2(23)(4)_4 \rightarrow (1)_2(324)_3 \nonumber \\
&&  (123)_1(4)_4 \rightarrow (3124) \rightarrow (1)_2(243)_3 \\
F^{(3)}_2 : && (123)_1(4)_4 \rightarrow (1)_3(23)(4)_4 \rightarrow (1)_3(324)_2 \nonumber \\
&&  (123)_1(4)_4 \rightarrow (3124) \rightarrow (1)_3(243)_2
\\
F^{(3)}_3 : && (12)_1(34)_4 \rightarrow (2134) \rightarrow (13)_2(24)_3 \nonumber \\
&&   (12)_1(34)_4 \rightarrow (1243) \rightarrow (13)_2(24)_3
\end{eqnarray}
We find the string diagram $S^{(3)}_1$ has a multiplicity of 1 with respect to $F^{(3)}_1$, the string diagram $S^{(3)}_2$ has a multiplicity of 1 with respect to $F^{(3)}_2$. The string diagram $S^{(3)}_3$ has the multiplicities of 1 with respect to both $F^{(3)}_1$ and $F^{(3)}_2$, and a multiplicity of 2 with $F^{(3)}_3$ from the previously missing diagram. We find that for  $S^{(3)}_1$  and  $S^{(3)}_2$, which represent the $T$ and $U$ channels of the $2\rightarrow 2$ scattering, the factorization relation holds, namely  
\begin{eqnarray}
 S^{(3)}_1 &=&  F^{(3)}_1 \nonumber \\
 S^{(3)}_2 &=&  F^{(3)}_2 
\end{eqnarray}
Now by including the new contribution $F^{(3)}_3$,  we find that the factorization also works for the $S$-channel process 
\begin{eqnarray}
S^{(3)}_3  &=&  F^{(3)}_1+F^{(3)}_2 + 2F^{(3)}_3
\end{eqnarray}
The total contribution to the correlator is now 
\begin{eqnarray}
\bra \bar{O}^{J_1} \bar{O}^{J_4}O^{J_2}O^{J_3}\ket =  F^{(3)}_1+F^{(3)}_2+ F^{(3)}_3
=  \frac{1}{2}( S^{(3)}_1+S^{(3)}_2 +S^{(3)}_3) 
\end{eqnarray}

\subsubsection{Case two: $\bra \bar{O}^{J_1} \bar{O}^{J_4}_{-m,m} O^{J_2}_0O^{J_3}_0 \ket$ ($J_1>J_2>J_3>J_4$)}

\begin{figure}
  \begin{center}
\includegraphics[width=6.5in]{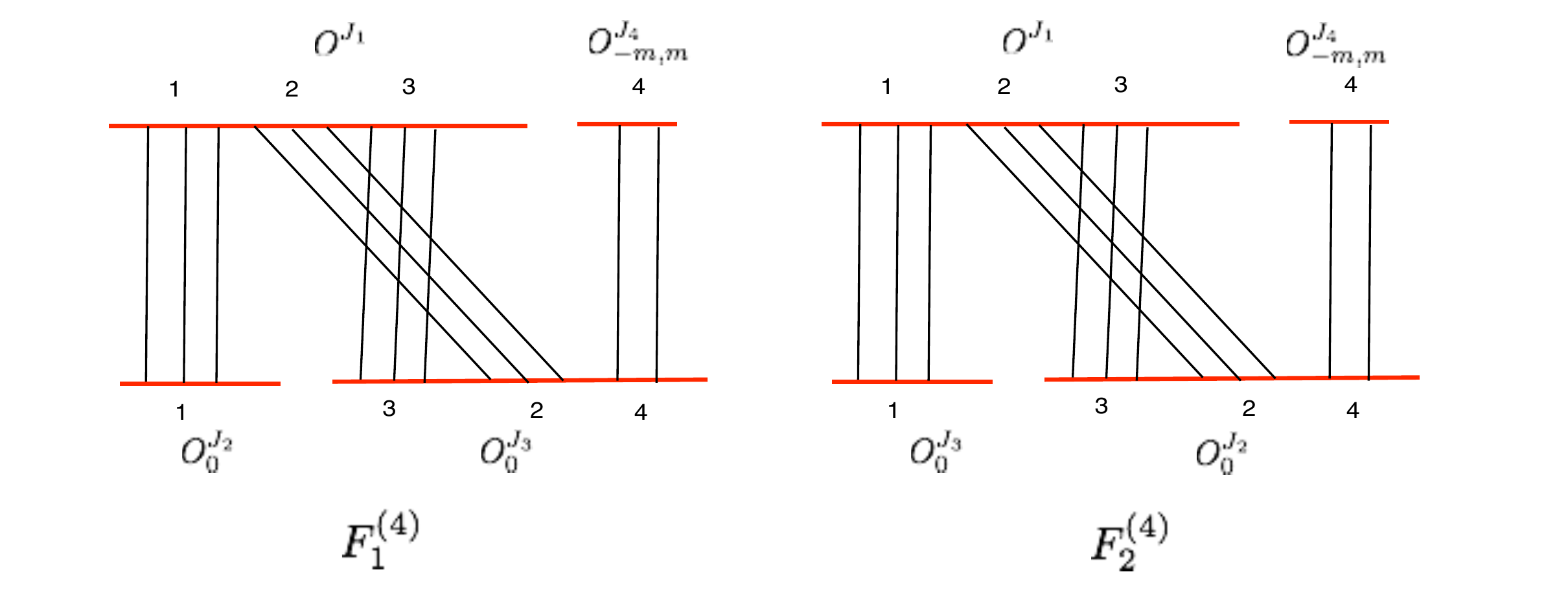} 
\end{center}
\caption{Some field theory diagrams for  $\bra \bar{O}^{J_1} \bar{O}^{J_4}_{-m,m} O^{J_2}_0O^{J_3}_0 \ket$ ($J_1>J_2>J_3>J_4$ ). These diagrams turn out to give vanishing contributions. }  \label{F4}
\end{figure}

The field theory diagrams in the previous paper \cite{Huang:2010} are depicted in Fig. \ref{F4}. In both cases it is impossible to put in the scalar insertions without violating planarity, so the contributions of these diagrams vanish
\begin{eqnarray} \label{F4eq}
F^{(4)}_1=0, ~~~~F^{(4)}_2=0. 
\end{eqnarray}

\begin{figure}
  \begin{center}
  \includegraphics[width=3.3in]{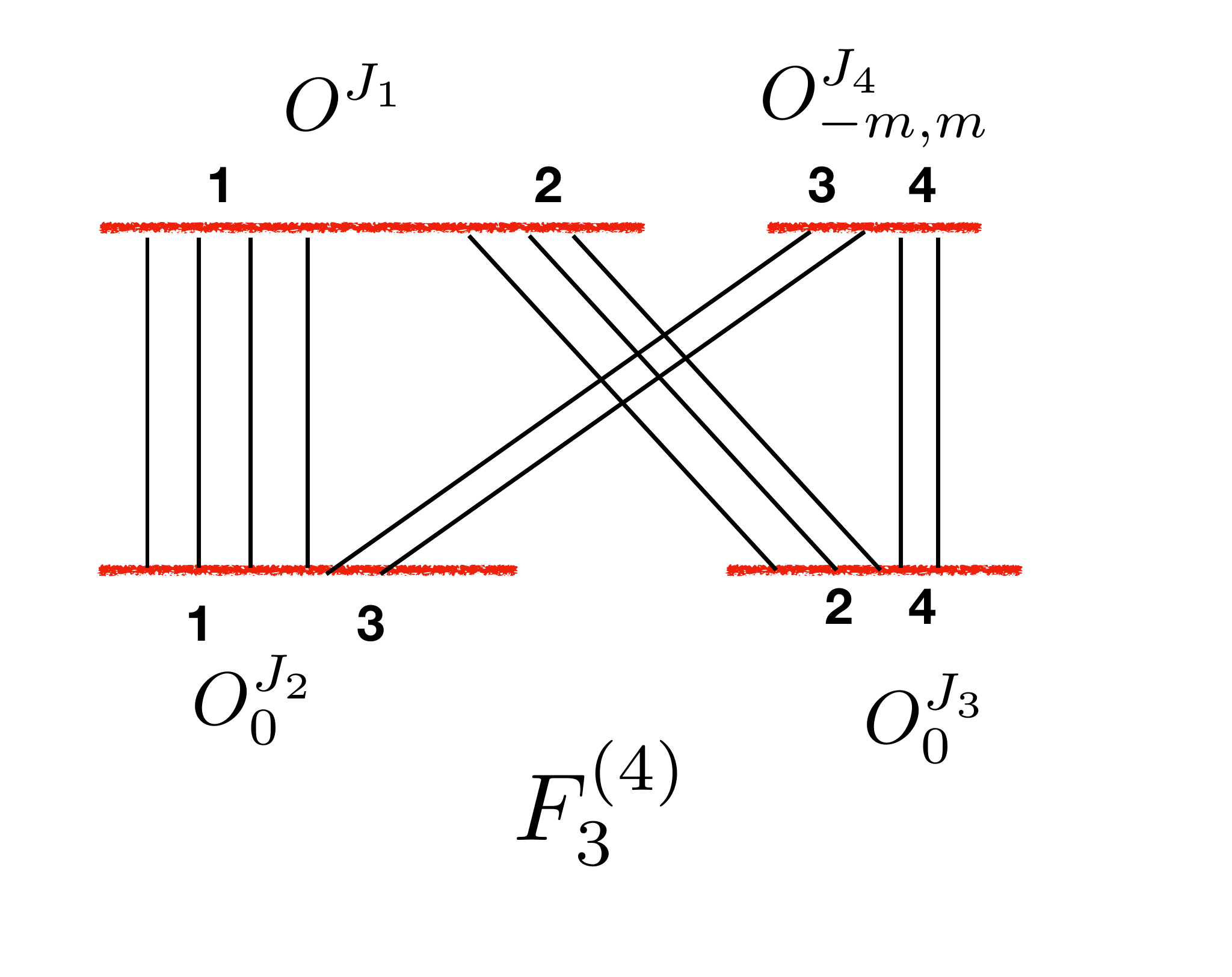} 
 \end{center}
\caption{The new diagram missed in our previous paper for   $\bra \bar{O}^{J_1} \bar{O}^{J_4}_{-m,m} O^{J_2}_0O^{J_3}_0 \ket$.}  \label{F4a}
\end{figure}

The missing diagram is depicted in Fig. \ref{F4a}, which is structurally the same as Fig. \ref{F3a}. But now it is possible to contract the scalar insertion in $O^{J_2}_0$ and $O^{J_3}_0$ without violating planarity. After carefully tracking the combinatorial factors, and separating the $Z$ fields in $O^{J_4}_{-m,m}$ into two parts, we can write the contributions as integral 
\begin{eqnarray} \label{F43eq}
F^{(4)}_3 &=&  \frac{g^2}{J} \sqrt{\frac{J_1}{J_4}}  \int_{0}^{x_4} dy \int_0^yd y_1 e^{2\pi i m \frac{y_1}{x_4}}  \int_y^{x_4} d y_2 e^{- 2\pi i m \frac{y_2}{x_4}} \nonumber \\ 
&=& - \frac{g^2}{J} \frac{(x_1)^{\frac{1}{2}} (x_4)^{\frac{5}{2}} }{2\pi^2 m^2} 
\end{eqnarray}

\begin{figure}
  \begin{center}
  \includegraphics[width=6.5in]{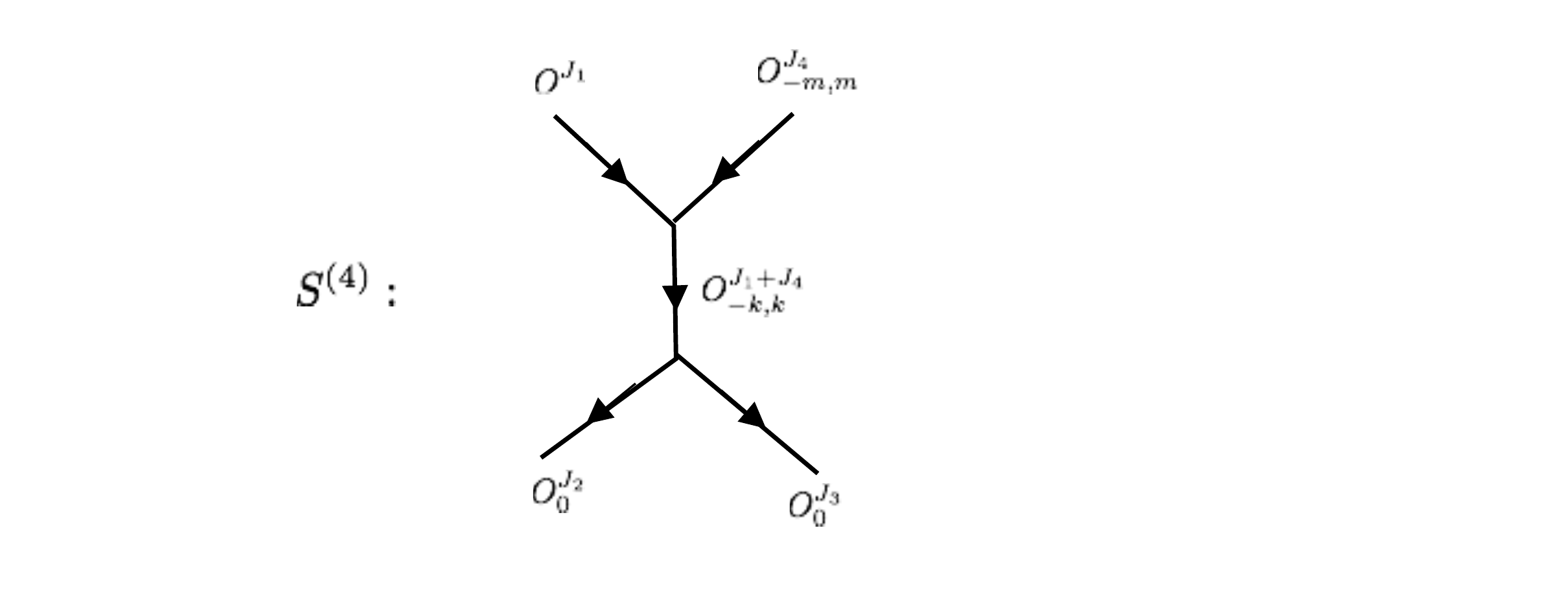} 
\end{center}
\caption{The string diagram for  $\bra \bar{O}^{J_1} \bar{O}^{J_4}_{-m,m} O^{J_2}_0O^{J_3}_0 \ket$ ($J_1>J_2>J_3>J_4$ ). This is the only nonvanishing $S$-channel diagram, which we denote  $S^{(4)}$. }  \label{S4}
\end{figure}

We look at the string diagrams. The longest operator is $O^{J_1}$, but it has no scalar insertion so it cannot decay to $O^{J_2}_0$ or $O^{J_3}_0$. So the $T$, $U$ channels are impossible and we are left only with the $S$-channel contribution $S^{(4)}$ depicted in Fig. \ref{S4}. The vanishing of the $T$, $U$ channels is consistent with the factorization rules since the field theory diagram contributions vanish (\ref{F4eq}). We can calculate the $S$-channel contribution  
\begin{eqnarray} 
S^{(4)} &=& \sum_{k=-\infty}^{+\infty}  \bra \bar{O}^{J_1} \bar{O}^{J_4}_{-m,m} O^{J_1+J_4}_{-k,k}\ket \bra \bar{O}^{J_1+J_4}_{-k,k} O^{J_2}_0O^{J_3}_0 \ket \nonumber \\
&=& -\frac{g^2}{J}  (x_1)^{\frac{1}{2}} (x_4)^{\frac{3}{2}}  \sum_{k=-\infty}^{+\infty} \frac{\sin^2(\pi k x_4)}{\pi^2 (kx_4-m)^2}  \frac{\sin^2(\pi k x_2)}{\pi^2 k^2}.  \label{S4eq} 
\end{eqnarray}
The infinite sum can be performed analytically using the useful summation formulas in the appendix in the previous paper \cite{Huang:2010}. To apply the summation formulas we note that it is convenient to use a trigonometry identity $4\sin^2(\beta_1) \sin^2(\beta_2) = 2\sin^2(\beta_1) +2\sin^2(\beta_2) - \sin^2(\beta_1+\beta_2) -\sin^2(\beta_1-\beta_2) $. Quite nicely, it turns out that the infinite sum in (\ref{S4eq}) is actually independent of $x_2$ as long as $x_2>x_4$ as assumed here.  With the new contributions, we check that indeed the $S$-channel factorization is satisfied,  
\begin{eqnarray}
\bra \bar{O}^{J_1} \bar{O}^{J_4}_{-m,m} O^{J_2}_0O^{J_3}_0 \ket = F^{(4)}_3  = \frac{1}{2} S^{(4)}
\end{eqnarray} 

 Here it is probably more illuminating to check the factorization rule with the integral form of the vertex (\ref{integralform}), as it was done in the previous paper for the torus two-point function \cite{Huang:2010}. The sum over intermediate states can be done by the Poisson resummation formula $\sum_{k=-\infty}^{\infty} e^{2\pi i kx}= \sum_{p=-\infty}^{+\infty}\delta(x-p) $. We can write the $S$-channel contribution as 
 \begin{eqnarray}
 S^{(4)} &=& \frac{g^2}{J} (\frac{x_1}{x_4})^{\frac{1}{2}} \int_0^{x_4} d y_1 e^{-2\pi i \frac{m}{x_4} y_1}  \int_0^{x_4} d y_2 e^{ 2\pi i \frac{m}{x_4} y_2}\nonumber \\ && \times \sum_{p=-\infty}^{\infty}\int_0^{x_2} d y_3 \int_{x_2}^1 d y_4 \delta(y_1-y_2+y_3-y_4-p). 
 \end{eqnarray} 
 We can analyze the integral. Since the final result is real, the contributions from $y_1>y_2$ and $y_2>y_1$ are the same. We consider for example $y_1>y_2$, then the sum over $p$ has nonvanishing contribution only for $p=0$. Since $x_4$ is the smallest, we have the integral $\int_{0}^{x_2} d y_3 \int_{x_2}^{1} d y_4  \delta(y_1-y_2+y_3-y_4) = y_1 - y_2 =\int_{y_2}^{y_1} d y $. Finally, we can rearrange the integral domain by $\int_0^{x_4} d y_1 \int_0^{y_1} d y_2 \int_{y_2}^{y_1} d y (\cdots)= \int_{0}^{x_4} d y \int_0^{y} d y_2 \int_y^{x_4} d y_1 (\cdots)$. This is exactly the integral formula in (\ref{F43eq}). Counting the double contributions from $y_1>y_2$ and $y_2>y_1$ we have the factorization rule $S^{(4)} = 2 F^{(4)}_3 $.

\subsubsection{Case three: $\bra \bar{O}^{J_1}_{-m,m} \bar{O}^{J_4} O^{J_2}_{-n,n}O^{J_3} \ket$  ($J_1>J_2>J_3>J_4$)}

\begin{figure}
  \begin{center}
  \includegraphics[width=5.5in]{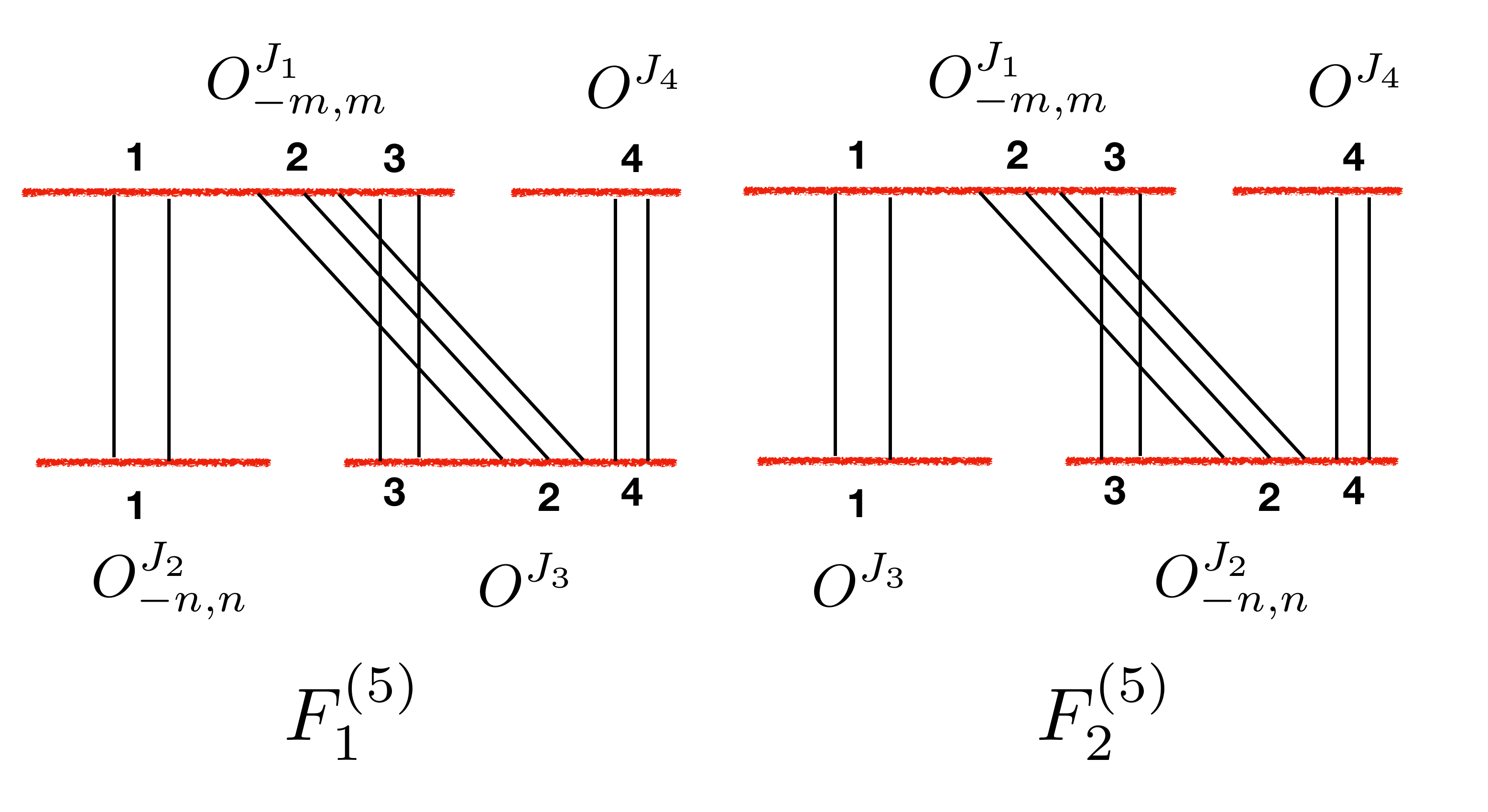} 
\end{center}
\caption{Some field theory diagrams for  $\bra \bar{O}^{J_1}_{-m,m} \bar{O}^{J_4} O^{J_2}_{-n,n}O^{J_3} \ket$. }  \label{F5}
\end{figure}

\begin{figure}
  \begin{center}
  \includegraphics[width=3.3in]{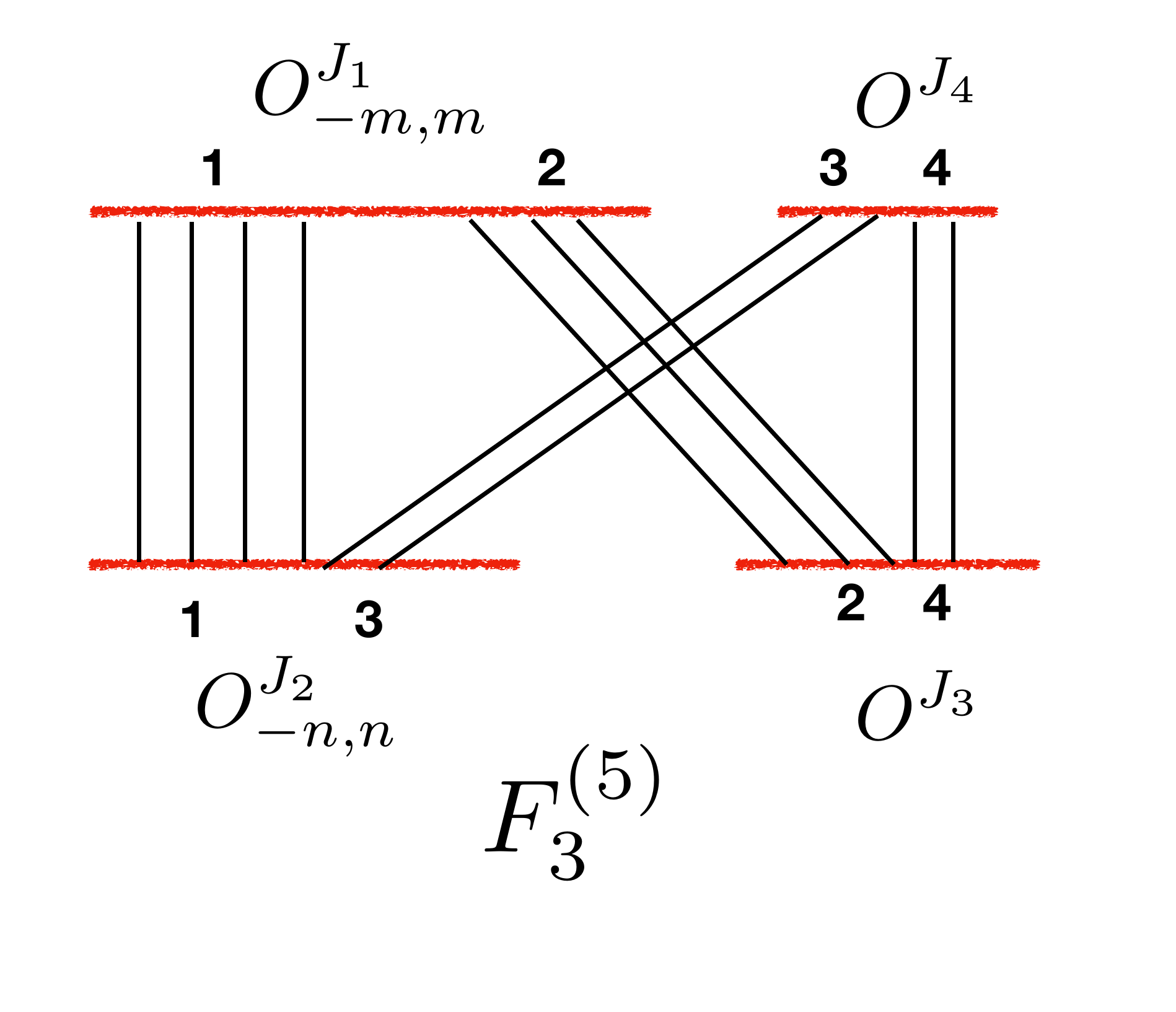} 
 \end{center}
\caption{The new diagram missed in our previous paper for  $\bra \bar{O}^{J_1}_{-m,m} \bar{O}^{J_4} O^{J_2}_{-n,n}O^{J_3} \ket$. }  \label{F5a}
\end{figure}

The field theory diagrams are depicted in Fig. \ref{F5} and \ref{F5a}. The two diagrams in Fig \ref{F5}  have been computed in the previous paper, and the results are 
\begin{eqnarray}
F^{(5)}_1 &=&  \frac{g^2}{J} (x_1)^{-\frac{1}{2}}x_2^{\frac{3}{2}} (x_3x_4)^{\frac{1}{2}} (x_1-x_2)\int_0^1 dy_1e^{-2\pi i (\frac{mx_2}{x_1}-n)y_1} \int_0^1 dy_2e^{2\pi i (\frac{mx_2}{x_1}-n)y_2} \nonumber \\
&=&  \frac{g^2}{J} (x_1x_2)^{\frac{3}{2}} (x_3x_4)^{\frac{1}{2}} (x_1-x_2)\frac{1-\cos(2m\pi\frac{x_2}{x_1})}{2\pi^2(mx_2-nx_1)^2},
\end{eqnarray}
\begin{eqnarray}
F^{(5)}_2 &=&  \frac{g^2}{J} (x_1x_2)^{-\frac{1}{2}} (x_3x_4)^{\frac{1}{2}} (x_1-x_3)^3 \int_0^1 dy  \times   \nonumber \\  &&
| (e^{- 2\pi i n\frac{x_1-x_3}{x_2}} \int_0^y dy_1+\int_y^1 dy_1) e ^{2\pi i (x_1-x_3)(\frac{m}{x_1}-\frac{n}{x_2})y_1} |^2  \nonumber \\
&=&  \frac{g^2}{J} \frac{(x_1x_2)^{\frac{3}{2}} (x_3x_4)^{\frac{1}{2}}}{2\pi^3(nx_1-mx_2)^3}\{ \pi(nx_1-mx_2)(x_1-x_3) 
[2-\cos(2m\pi\frac{x_3}{x_1})-\cos(2n\pi\frac{x_4}{x_2})] \nonumber \\ && 
+x_1x_2[\sin(2m\pi \frac{x_3}{x_1})-\sin(2n\pi\frac{x_4}{x_2})-\sin(2\pi\frac{(nx_1-mx_2)(x_1-x_3)}{x_1x_2})] \}
\label{F52}
\end{eqnarray}

\begin{figure}
  \begin{center}
  \includegraphics[width=6.5in]{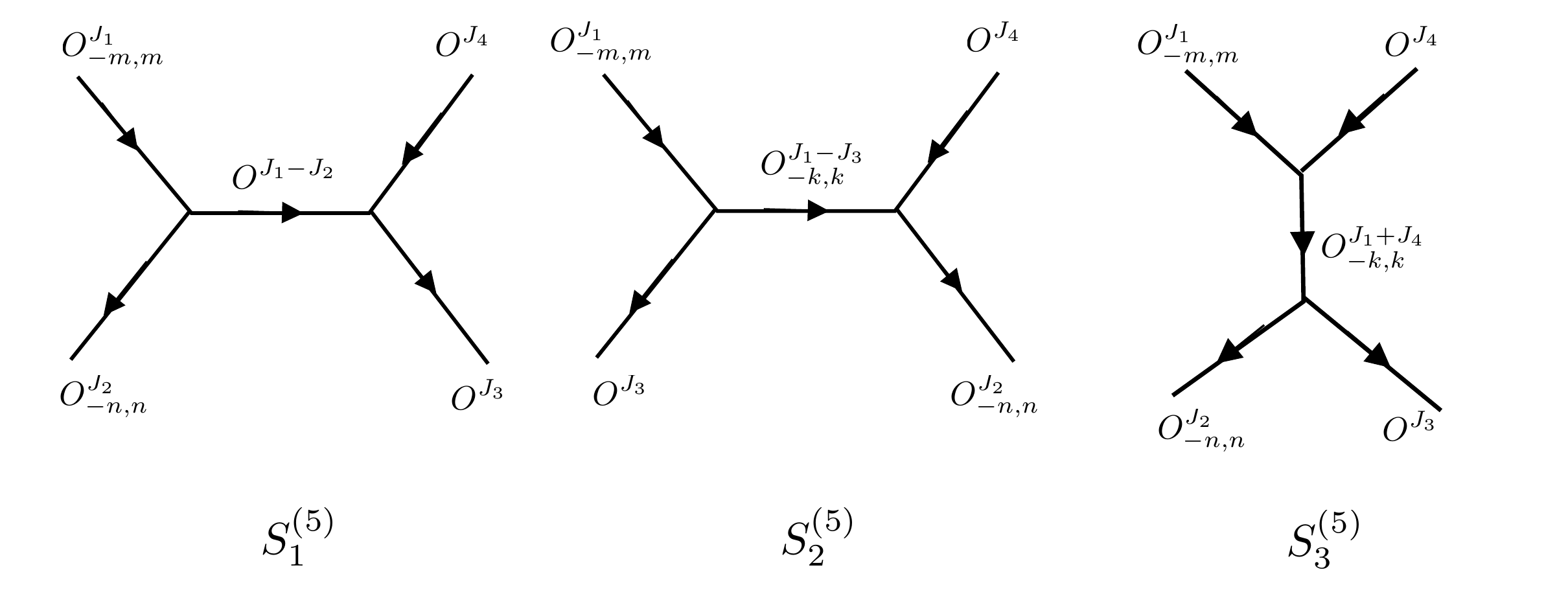} 
\end{center}
\caption{The string diagrams for  $\bra \bar{O}^{J_1}_{-m,m} \bar{O}^{J_4} O^{J_2}_{-n,n}O^{J_3} \ket$ ($J_1>J_2>J_3>J_4$ ). We denote the contributions of the two diagrams $S^{(5)}_1$,  $S^{(5)}_2$, $S^{(5)}_3$ respectively. }  \label{S5}
\end{figure}

The string diagrams are depicted in Fig. \ref{S5}. In the previous paper we have computed  the $T,U$ channels contributions denoted by $S^{(5)}_1$, $S^{(5)}_2$, and verified the factorization relation.  
\begin{eqnarray}
S^{(5)}_1= \bra \bar{O}^{J_1}_{-m,m} O^{J_2}_{-n,n}O^{J_1-J_2}\ket \bra \bar{O}^{J_1-J_2}\bar{O}^{J_4} O^{J_3} \ket
=F^{(5)}_1,
\end{eqnarray}
\begin{eqnarray}
S^{(5)}_2 = \sum_{k=-\infty}^{+\infty} \bra \bar{O}^{J_1}_{-m,m} O^{J_3} O^{J_1-J_3}_{-k,k}\ket \bra \bar{O}^{J_1-J_3}_{-k,k}\bar{O}^{J_4} O^{J_2}_{-n,n} \ket = F^{(5)}_2
\end{eqnarray}

We calculate the new diagram in Fig. \ref{F5a}. The two oscillator modes are located at the left side of the diagram, 
\begin{eqnarray}
F^{(5)}_3 &=& \frac{g^2}{J} (\frac{x_3x_4}{x_1x_2})^{\frac{1}{2}} \int_0^{x_4} dy~ |\int_0^{x_2-y} e^{-2\pi i (\frac{m}{x_1} - \frac{n}{x_2}) y_1 } d y_1|^2 
\nonumber \\ 
&=&  \frac{g^2}{J} \frac{(x_1x_2)^{\frac{3}{2}} (x_3x_4)^{\frac{1}{2}}}{4\pi^3(nx_1-mx_2)^3} [2\pi (nx_1-mx_2)x_4 
+x_1x_2\sin(2m\pi\frac{x_2}{x_1}) \nonumber \\ && + x_1x_2 \sin(2\pi\frac{(nx_1-mx_2)(x_2-x_4)}{x_1x_2}) ]. 
\end{eqnarray}

It is quite involved to use the integral form and Poisson resummation formula to check the $S$-channel factorization. It is more straightforward to perform the infinite sum using the formulas in the appendix of our previous paper \cite{Huang:2010}. We check indeed that 
\begin{eqnarray}
S^{(5)}_2 = \sum_{k=-\infty}^{+\infty} \bra \bar{O}^{J_1}_{-m,m} \bar{O}^{J_4}O^{J}_{-k,k}\ket \bra \bar{O}^{J}_{-k,k} O^{J_2}_{-n,n}  O^{J_3}  \ket = F^{(5)}_1+F^{(5)}_2 + 2F^{(5)}_3 .
\end{eqnarray}

The total contribution to the correlator is 
\begin{eqnarray}
\bra \bar{O}^{J_1}_{-m,m} \bar{O}^{J_4} O^{J_2}_{-n,n}O^{J_3} \ket =F^{(5)}_1+F^{(5)}_2+F^{(5)}_3=\frac{1}{2} (S^{(5)}_1+S^{(5)}_2+S^{(5)}_3) .
\end{eqnarray}

. 

\section{Some observations about higher genus correlators} 
\label{sec3}

In quantum mechanics, for a  initial state $|\psi\ket$ with the proper normalization ${\bra \psi|\psi\ket=1}$, we may make a measurement about a physical observable such as the energy. It is well known that the eigenstates of the  physical observable form a complete orthonormal basis. Summing the probability of all possible final states is simply the unity probability $\sum_{n} |\bra \psi|n\ket|^2=1$.

Analogously, we interpret the two single-string correlator $\bra \bar{O}^{J}_{-m,m}  O^{J}_{-n,n} \ket_{h}$ of genus $h$ as the physical $h$-loop probability amplitude  of preparing the initial as $O^{J}_{-m,m}$ and observing the final state $O^{J}_{-n,n}$. We note that since the spacetime is infinitely curved, a hypothetical observer is performing the gedanken experiment of measurement without an ambient spacetime. First we consider the torus two-point function 
\begin{eqnarray}\label{F6torus}
 && \bra \bar{O}_{-m,m}^J O_{-n,n}^J \ket_{\textrm{torus}}  \\ 
&=& \left\{
\begin{array}{cl}
\frac{g^2}{24},    &   m=n=0;   \\
0,              &  m=0, n\neq0,   \\
&  \textrm{or}~n=0, m\neq0; \\
g^2(\frac{1}{60} - \frac{1}{24 \pi^2 m^2} + \frac{7}{16 \pi^4 m^4}),   &  m=n\neq0; \\
\frac{g^2}{16\pi^2m^2} ( \frac{1}{3}+\frac{35}{8\pi^2m^2}),  &  m=-n\neq0;    \nonumber \\
\frac{g^2}{4\pi ^{2}(m-n)^2} ( \frac{1}{3}+\frac{1}{\pi
^2n^2}+\frac{1}{\pi ^2m^2}-\frac{3}{2\pi ^2mn}-\frac{1}{2\pi
^2(m-n)^2}) & \textrm{all~other~cases} 
\end{array}
\right.
\end{eqnarray}
We find that the sum of the probability amplitude over final states is actually quite simple and independent of the initial state mode
\begin{eqnarray} \label{sumover} 
\sum_{n=-\infty}^{\infty} \bra \bar{O}_{-m,m}^J O_{-n,n}^J \ket_{\textrm{torus}} = \frac{g^2}{24}. 
\end{eqnarray}
It is easy to check the infinite sum for some fixed small integers $m$. The result can be better derived using the integral formula 
\begin{eqnarray} \label{torusintegral}
&& \bra \bar{O}_{-m,m}^J O_{-n,n}^J \ket_{\textrm{torus}}   \\ 
&=& g^2 \int_{0}^1 d x_1 dx_2 dx_3 dx_4 \delta(x_1+x_2+x_3+x_4-1)
[\int_0^{x_1} dy_1 e^{-2\pi i (m-n) y_1}]    \nonumber \\
&& \cdot  [\int_0^{x_1} dy_2 e^{2\pi i (m-n) y_2} +  e^{2\pi im(x_3+x_4) } \int_{x_1}^{x_1+x_2} dy_2 e^{2\pi i (m-n) y_2} 
\nonumber \\ &&
+e^{2\pi i m(x_4-x_2) } \int_{x_1+x_2}^{1-x_4} dy_2 e^{2\pi i (m-n) y_2} 
+e^{-2\pi i m(x_2+x_3) } \int_{1-x_4}^{1} dy_2 e^{2\pi i (m-n) y_2} ].  \nonumber
\end{eqnarray}
Summing over $n$ with the Poisson resummation formula gives a delta function, which only has nonvanishing contribution for the first of the four terms in the last factor of the integrand. The delta function also ensures that the dependence on initial state  mode $m$ cancels out. The final integral is simply $\int_{0}^1 d x_1 dx_2 dx_3 dx_4 \delta(x_1+x_2+x_3+x_4-1) x_1=\frac{1}{24}$, proving the formula (\ref{sumover}). 

An alternative derivation is to use factorization formula 
\begin{eqnarray}
\bra \bar{O}_{-m,m}^J O_{-n,n}^J \ket_{\textrm{torus}} &=& \frac{1}{2} (\sum_{J_1=1}^{J-1}\sum_{k=-\infty}^{+\infty} \langle
\bar{O}^J_{-m,m} O^{J_1}_{-k,k} O^{J-J_1} \rangle \langle
\bar{O}^{J_1}_{-k,k} \bar{O}^{J-J_1} O^J_{-n,n}
\rangle \nonumber \\ &&+
 \sum_{J_1=1}^{J-1}\langle \bar{O}^J_{-m,m} O^{J_1}_0 O^{J-J_1}_0
\rangle \langle \bar{O}^{J_1}_0 \bar{O}^{J-J_1}_0
O^J_{-n,n} \rangle   ) \label{factorizationtorus}
\end{eqnarray}
We can perform the calculations using the integral form of the vertex (\ref{integralform}) and sum over final states.  Again using the Poisson resummation formula, we derive the formula (\ref{sumover}). Here the second term in (\ref{factorizationtorus}) vanishes with summing over $n$, and the contribution comes entirely from only the first term.

Formula (\ref{sumover}) supports the interpretation of $\bra \bar{O}_{-m,m}^J O_{-n,n}^J \ket _{\textrm{torus}} $ as a kind of observable probability amplitude, as the proper normalization to unity is independent of initial string mode $m$. It might seem a little strange that unlike usual quantum mechanics, we do not need to take the norm square of the amplitude. The underlying reason becomes clear later. For now this is not a problem since one can easily show that  $\bra \bar{O}_{-m,m}^J O_{-n,n}^J \ket _{\textrm{torus}} $ is already real and non-negative. Furthermore, summing instead the square of the torus two-point function over final modes $n$ would be dependent on initial mode $m$, and would not give a nice formula like (\ref{sumover}).

One may wonder what happens to multistrings in the context of our probability interpretation. For example, the planar three-point amplitude $\langle\bar{O}^J_{-m,m}O^{J_1}_{0}O^{J_2}_{0}\rangle$ for $m\neq 0$ seems to be negative. However, this is not really a problem because they actually vanish in the BMN limit where $J, N\rightarrow \infty$, noticing the factor of $\sqrt{J}$  in all three-point functions in (\ref{planar1}). Only the (higher genus) two-point functions remains finite in this strict  BMN limit. So we argue that the single string states form a complete Hilbert space by themselves, and we can regard the multistrings as virtual states only appearing in the intermediate steps of a physical process. In this sense an amplitude with external multistring states such as the $2\rightarrow 2$ amplitudes discussed in Sec \ref{sec2} is a virtual amplitude. The virtual amplitudes are still useful because they can make a finite contribution to the physical two-point amplitudes in a string loop diagram calculation due to also a large number $J$ ways to split a single string into two strings, as e.g. in the factorization formula (\ref{factorizationtorus}).  In this way, the (vanishing) contributions of general multistrings have been already accounted for when we study two-point functions, due to formulas like (\ref{factorizationtorus}). It would be redundant to include them as part of the physical Hilbert space. 

This is a tricky point so we provide further clarification. It may be helpful to forget about the factorization formula for a moment. The factorization formulas provide an alternative derivation of some technical facts, but are not otherwise essential for this section. Let us consider free gauge theory and take the ``strict" BMN limit where $J, N\rightarrow \infty$, with $g=\frac{J^2}{N}$ being finite. Then the general $n$-point correlators for $n>2$ vanish as they are always suppressed by positive power of $J$. Two-point functions of BMN operators are the only relevant finite physical quantities to be considered. In this strict limit a single string cannot really decay into multistrings at all, which is the reason that they are called virtual. So the multistrings should not pose a problem here since we regard the results in this section as only valid in the strict BMN limit.

In quantum mechanics we can multiply the quantum state by a complex phase factor and do not change the underlying physical quantum state. This seems to pose a problem for our probability interpretation. For example, if we change the initial state BMN operator by a minus sign, then the two-point function would become negative, and inconsistent with a probability interpretation. Our strategy here is to first work with the BMN states, and fix a uniform phase factor for the operators, avoiding the problem for the moment. Indeed the BMN operators form a preferred natural basis for physical states. If we deform away from the infinite curvature limit in the pp-wave background, corresponding to turning on gauge interactions on the field theory side, the mass degeneracy of string excited states is broken. The BMN operators were originally proposed to correspond to the string mass eigenstates, and their conformal dimensions in the planar limit compute the free string mass spectrum \cite{BMN}.  In this sense the BMN strings are the ``on-shell" states that appear as external states in usual quantum field theory calculations of scattering amplitudes. We argue that this basis is preferred even in the infinite curvature limit where the string mass spectrum is degenerate. Of course, the fundamental principle of quantum mechanics implies that there should be  complex linear superpositions of BMN states. We explain later how to deal with them. 

In quantum mechanics, an observer performs measurement with respect to an orthogonal basis of states. Here since there appears to be nontrivial overlap between BMN strings, one may wonder whether they form an orthogonal basis for measurement. To understand this point, we can compare again with scattering amplitudes in quantum field theory, where the external states are mass eigenstates of free theory. The interactions are turned off with the external states propagating asymptotically away to spacetime infinity. Likewise, we argue here that when a hypothetical observer measures the external BMN states, the string interactions are turned off. The BMN operators are certainly orthogonal for zero string coupling $g=0$ where we have only planar contributions, so are consistent for quantum measurement. While in quantum field theory we put the interaction vertices in the middle of initial and final states to compute the scattering amplitudes, in our case the BMN strings self-interact by splitting and rejoining when we turn on the string coupling $g>0$.

For general higher genus $h$, the field theory has the $ \frac{(4h-1)!!}{2h+1}$ cyclically different diagrams  \cite{HZ}. Similar to the genus one case, using the field theory integrals, one can show the contribution of each diagram summing over all final states is $ \frac{g^{2h}}{(4h)!}$. So the formula is 
\begin{eqnarray}
\sum_{n=-\infty}^{\infty} \bra \bar{O}_{-m,m}^J O_{-n,n}^J \ket_{h} = \frac{(4h-1)!!}{(2h+1)(4h)!} g^{2h}.
\end{eqnarray}
This is the coefficient of the series expansion of the all-genus formula for the vacuum correlator $\bra \bar{O}^J O^J \ket_{\textrm{all genera}} = \frac{2\sinh(g/2)}{g}$. 

An immediate consequence of the probability interpretation is that $\bra \bar{O}_{-m,m}^J O_{-n,n}^J \ket_{h}$ must always be non-negative. This fact can be understood from two perspectives. From the field theory perspective, we are doing integrals of two oscillator modes along the string. For example, in the formula for torus integral (\ref{torusintegral}), for one oscillator mode we only need to do the integral for one of the four segments, due to cyclicity. We can also write it in a more complicated but symmetric form that treats both oscillator modes equally. Then the integrand is positively proportional to the product of two complex conjugate parts, so is always non-negative. On the other hand, from string theory perspective, up to two scalar modes, the only negative cubic vertex is   $\langle\bar{O}^J_{-m,m}O^{J_1}_{0}O^{J_2}_{0}\rangle$ for $m\neq 0$. This type of vertex separates two string oscillator modes into two strings, and must always appear in even numbers in a string diagram of $\bra \bar{O}_{-m,m}^J O_{-n,n}^J \ket_{h}$. Here we need to be a little careful to discuss some special cases since $\langle\bar{O}^J_{0,0}O^{J_1}_{0}O^{J_2}_{0}\rangle$ is positive. If both $m,n$ are not $0$, or $m=n=0$, the negative vertex always appear in pairs and the string diagram contribution is positive. In the special case one of $m,n$ is $0$ and the other is not, some string diagram contributions can be negative, but the overall contribution actually vanishes e.g. $\bra \bar{O}_{-m,m}^J O_{0,0}^J \ket_{h} = 0$ for $m\neq 0$, and is still of course consistent with a probability interpretation. 

In the field theory computations, we see that the BMN correlators are secretly accounting for the norm square in the usual quantum probability in the integrand. The underlying reason is the close string level matching condition which requires that the two string modes in a BMN operator have opposite sign. 

Putting together the results, we can write the total probability $p_{m,n}\geq 0$ of preparing an initial single string state $O_{-m,m}$, then observing a final single string state $O_{-n,n}$, including all string loop contributions, as 
\begin{eqnarray} \label{pmatrix}
p_{m,n} = \frac{g}{2\sinh(g/2) } \sum_{h=0}^{\infty} \bra \bar{O}_{-m,m}^J O_{-n,n}^J \ket_{h} .
\end{eqnarray} 
This is properly normalized by the vacuum correlator so that $\sum_{n=-\infty}^{\infty} p_{m,n} =1$ for any initial mode $m$.

We can now discuss  our proposal more precisely in terms of  the usual formulation of quantum mechanics. We denote the orthonormal BMN states of free string theory by $|n\ket$.  Let us assume that the transition amplitude between BMN states can be described by a unitary operator $e^{i\hat{H}(g)}$, where  $\hat{H}(g)$ is a Hermitian operator corresponding to the time integral of Hamiltonian in a usual quantum mechanics system. Here $\hat{H}(0)=0$ for free string theory and the operator $\hat{H}(g)$ models string interactions at finite coupling $g$.  Our probability interpretation implies that the matrix element $p_{m,n}$ in (\ref{pmatrix}) does not correspond naively to the usual transition amplitude $\bra m | e^{i\hat{H}(g)} |n\ket$, but rather to its norm square $|\bra m | e^{i\hat{H} (g)} |n\ket|^2$. This is already strongly supported by the same normalization relation $\sum_{n=-\infty}^{\infty} p_{m,n} =\sum_{n=-\infty}^{\infty} |\bra m | e^{i\hat{H} (g)} |n\ket|^2 = 1$. 

We should note that most studies of AdS/CFT have been focused on supergravity approximation, BPS operators in CFT. The precise dictionary of AdS/CFT has not been much explored beyond supergravity. The standard holographic dictionary \cite{Gubser, Witten} may not be directly applicable here since the geometry is no longer AdS space and we are dealing with stringy modes.  It is helpful to clarify the relation to the operator/state correspondence, which is a well-known property of any CFT relating the two-point function to an overlap of the corresponding states in CFT.  We also expect the correspondence of BMN operators with holographic dual string states, at least for free string theory, while string interactions might better be described by a unitary operator. \footnote{It would be extremely awkward to directly quantize an interacting string theory, giving rise to string states that are not orthogonal. We do not know any such example. } However the dual string states are not exactly the same as the CFT states from the operator/state correspondence. As such, in our context, the two-point function is not necessarily an overlap of  string states.  It may be an overlap of CFT states according to the operator/state correspondence. However this is an entirely different matter and we are not considering this aspect at all. Our proposal might seem surprising, but we are not aware of any conflict with the well-established aspects of AdS/CFT correspondence. In fact we think this is an interesting new entry in the holographic dictionary. 

In general, a purely real perturbative series is not expected to be unitary. Here, in particular, we show that naively taking  $p_{m,n}$ in (\ref{pmatrix}) as the transition amplitude $\bra m | e^{i\hat{H}(g)} |n\ket$ would indeed be inconsistent with unitarity. We allow the generous possibility that the BMN operator $O_{-m,m}^J$  may be normalized by a function $f_m(g)$, depending on both mode number and string coupling. The unitarity condition is then 
\begin{eqnarray} \label{naiveunitarity}
\sum_{n=-\infty}^{\infty} f_m( g)^* f_{m^{\prime}}( g) |f_n( g)|^2 p_{m,n}  p^*_{m^{\prime},n}  = \delta_{m,m^{\prime}} . 
\end{eqnarray}
Consider the simple case $m=m^{\prime}$ and $g=0$ we deduce $|f_m(0)|=1$ for any mode $m$. Next we consider a nontrivial case $m\neq m^{\prime}$ and both nonzero. Expanding for small $g$, the leading term of the left-hand side of (\ref{naiveunitarity}) is of order $g^2$ from two contributions $n=m, m^{\prime}$ in the summation 
\begin{eqnarray} 
&& \sum_{n=-\infty}^{\infty} f_m( g)^* f_{m^{\prime}}( g) |f_n( g)|^2 p_{m,n}  p^*_{m^{\prime},n}    \nonumber \\
&=&2 f_m( 0)^* f_{m^{\prime}}( 0)   \bra \bar{O}_{-m,m}^J O_{-m^{\prime}, m^{\prime}}^J \ket_{\textrm{torus}}  +\mathcal{O}(g^3) . 
\end{eqnarray}
This is clearly nonzero, violating unitarity.  So we conclude if we believe the free single-string states form a complete Hilbert space $\sum_n |n\ket \bra n|=1$, that the naive proposal is basically ruled out.

Coming back to the correct proposal, by introducing some real phase angles, we can write the transition amplitude as 
\begin{eqnarray} \label{transition} 
\bra m | e^{i\hat{H}(g)} |n\ket = e^{i\theta_{m,n}(g)} \sqrt{p_{m,n}}. 
\end{eqnarray}
For $g=0$ this is the identity matrix $\delta_{m,n}$, so the phase angle $\theta_{m,m}(0)=0$. For $m\neq n$, the phase angle  $\theta_{m,n}(0)$ is not determined this way since in this case $p_{m,n}$ already vanishes for $g=0$. Another special case is that for $m\neq 0$, the matrix elements $p_{m,0}$ and $p_{0,m}$ vanish for any coupling $g$, so the phase angles $\theta_{m,0}(g), \theta_{0, m}(g)$ are actually redundant.  The unitarity condition is 
\begin{eqnarray} \label{unitarity}
\sum_{n=-\infty}^{+\infty} e^{i [ \theta_{m,n}(g) - \theta_{m^{\prime},n}(g) ]} \sqrt{p_{m,n} p_{m^{\prime},n}} = \delta_{m,m^{\prime}}. 
\end{eqnarray}
For $m=m^{\prime}$ this is already satisfied, while the cases of $m\neq m^{\prime}$ may provide some constraints for the phase angles. For the special case where one of $m,m^{\prime}$ is zero and the other is not, the unitary condition (\ref{unitarity}) is automatically satisfied without any constraint for the phase angles. Again we consider the nontrivial case of $m\neq m^{\prime}$ and both nonzero. Expanding the left-hand side of (\ref{unitarity}) for small $g$, we see the leading order term is now of order $g$ with two contributions also from $n=m, m^{\prime}$ in the summation 
\begin{eqnarray}
&& \sum_{n=-\infty}^{+\infty} e^{i [ \theta_{m,n}(g) - \theta_{m^{\prime},n}(g) ]} \sqrt{p_{m,n} p_{m^{\prime},n}}  \nonumber \\ &=& [e^{i  \theta_{m,m^{\prime} }(0)} + e^{- i  \theta_{m^{\prime},m }(0) }] \sqrt{ \bra \bar{O}_{-m,m}^J O_{-m^{\prime}, m^{\prime}}^J \ket_{\textrm{torus}} } +\mathcal{O}(g^2) . 
\end{eqnarray}
Since the torus two-point function is nonzero, we arrive at an interesting relation imposed by unitarity that $ \theta_{m,m^{\prime} }(0) +  \theta_{m^{\prime} ,m}(0) =\pi$, up to an integer multiple of $2\pi$, for nonzero $m\neq m^{\prime}$. So the phase angles cannot be trivially all set to $0$ even for free string theory, due to the consistency of string interactions. With these extra phase angles, we see that we are able to preserve the unitarity condition violated by the naive proposal.

To illustrate the power of unitarity, let us consider an analogous simpler situation of a Hilbert space of finite dimension $D$, and count the degrees of freedom. In this case there are $D^2$ real phase angles, and the unitarity conditions provide $\frac{D(D-1)}{2}$ complex equations. It might seem that generically we can solve the unitarity equations with $D$ remaining free real parameters. However some of the constrains are not independent and we actually have $2D-1$ remaining free real parameters. For example, for $D=2$ it is simple to parametrize a general $2\times 2$ unitary matrix
\begin{eqnarray}
U = e^{i\phi} \begin{pmatrix}
  e^{i\phi_1} \cos\alpha  & e^{i\phi_2} \sin\alpha  \\ 
  -e^{-i\phi_2} \sin\alpha & e^{-i\phi_1} \cos\alpha
 \end{pmatrix}. 
\end{eqnarray}
We see that after fixing the norms of the matrix elements, we still have 3 free phase angles. These $2D-1$ free real parameters can be easily seen from the unitarity condition (\ref{unitarity}). Once a particular solution $\theta_{m,n}$ is found, we can always shift $\theta_{m,n}\rightarrow \theta_{m,n}+\theta_m$ or $\theta_{m,n}\rightarrow \theta_{m,n}+\theta_n^\prime$, which are still solutions of the unitarity equations. There are $2D-1$ free parameters since the two types of shifts are the same if all the shifted angles $\theta_n, \theta_n^\prime$ are the same. Another perspective is to note that the norm square sums of each column and each row of a unitary matrix are always 1. To construct a general unitary matrix, we can first set the norms of the elements in a $(D-1)\times (D-1)$ block of the matrix to $(D-1)^2$ free real parameters, and the remaining matrix element norm would then be fixed. There are $D^2$ free real parameters for a general unitary matrix, so there must be $D^2-(D-1)^2 = 2D-1$ remaining free parameters for the phase angles.

We can also discuss some physical freedom of gauge choices. First, there is a freedom to rotate the base states by a phase factor $|n\ket \rightarrow e^{i \theta_n} |n\ket$, so the phase angles in transition amplitude (\ref{transition}) are shifted $\theta_{m,n} \rightarrow  \theta_{m,n} + \theta_n - \theta_m$. There are $D-1$ free parameters for such choices since  $\theta_{m,n}$ is unchanged if all $ \theta_n$'s are the same. Secondly, the overall phase corresponds to a shift of the operator $\hat{H}$ by a real number, which is a zero-point energy, usually considered physically unobservable without coupling the quantum mechanics to gravity.  We note that in our case since the zero mode decouples from the nonzero modes, there are two free choices of the ``zero-point energy,"  the phase angle $\theta_{0,0}(g)$ and the overall phase for the nonzero modes. For free string theory we have already made such choices by setting $\hat{H}(0)=0$. We can now further simply choose $\theta_{0,0}(g) = 0$ for any string coupling $g$, and focus on the nontrivial nonzero modes. A simple example of fixing all gauge choices for the nonzero mode block is to choose $\theta_{1,m}(g) = 0$ for all $m$'s and any coupling $g$.

These physical gauge choices are included in the $2D-1$ free parameters in phase angles from the shifts $\theta_{m,n}\rightarrow \theta_{m,n}+\theta_m$ or $\theta_{m,n}\rightarrow \theta_{m,n}+\theta_n^\prime$. So for a generic D-dimensional unitary matrix, we still have $D-1$ remaining free real parameters after fixing the above physical gauge choices. For the actual case of infinite dimensional Hilbert space, these infinitely many free parameters in the phase angles are thus not in principle determined in our setting. It would be interesting to study whether they can be determined by other methods.

It would be interesting to fully explicitly compute these phase angles $\theta_{m,n}(g)$ for general coupling $g$. In any case, now we can, in principle, follow the usual rule of quantum mechanics to compute the transition amplitudes between complex linear superpositions of BMN states with Eq. (\ref{transition}). We note that we cannot compute  by naively putting linear combinations of BMN operators in the correlators in (\ref{pmatrix}). Of course, this does not violate the fundamental principle of linear superposition in quantum mechanics, as we now understand that the two-point functions in (\ref{pmatrix}) do not directly correspond to the inner product of the underlying quantum system. This simply means we need to work a little more carefully instead with the right formula (\ref{transition}). Our earlier strategy of fixing a uniform phase for the BMN operators is also now justified.

In usual quantum mechanics, the diagonalization of the transition amplitude gives rise to the energy eigenstates. Since here I restrict myself to free gauge theory with only string interactions, the BMN states already have completely degenerate mass. Here the quantum transition amplitude (\ref{transition}) is more like a S-matrix where the incoming and outgoing states have the same energy. So the eigenstates that diagonalize the transition amplitude, i.e. eigenstates of the Hermitian operator $\hat{H}(g)$, are probably not the conventional energy eigenstates. It will be interesting to explore the physical interpretations of such eigenstates in the future.

Similar results are also true for BMN operators with more oscillator modes. For example, the BMN operators with three scalar excitation modes and orthonormal at planar level are  
\begin{eqnarray}  
O^{J}_{(m_1,m_2,m_3)} = \frac{1}{\sqrt{N^{J+2}}J} \sum_{l_1, l_2=0}^{J}  e^{\frac{2\pi im_2l_1}{J}} e^{\frac{2\pi im_3l_2}{J}}  \Tr(\phi^1 Z^{l_1} \phi^2 Z^{l_2-l_1} \phi^3 Z^{J-l_2}),  \label{operator3}
\end{eqnarray}
where the three scalar modes satisfy the close string level matching condition $m_1+m_2+m_3=0$. The factorization formula was studied in our previous paper \cite{Huang:2010}. We can show similarly that the sum over final states is independent of the initial state modes, 
\begin{eqnarray} \label{sumover3} 
\sum_{n_1+n_2+n_3=0}^{\infty} \bra \bar{O}^{J}_{(m_1,m_2,m_3)} O^{J}_{(n_1,n_2,n_3)}  \ket_{\textrm{torus}} = \frac{g^2}{24}. 
\end{eqnarray}
We also check that $ \bra \bar{O}^{J}_{(m_1,m_2,m_3)} O^{J}_{(n_1,n_2,n_3)}  \ket_{\textrm{torus}} \geq 0$ is always true including various degenerate cases, consistent with the physical probability interpretation.

Recently  Erbin et al. considered the two-point string amplitudes, and showed that they are not as trivial as previously thought \cite{Erbin:2019uiz}. These authors consider the conventional flat space, while we consider the infinitely curved pp-wave background. Despite the different settings, we see the development as another supporting evidence that the higher genus two-point amplitudes discussed here should have very relevant physical interpretations.

\vspace{0.2in} {\leftline {\bf Acknowledgments}}

This work was supported by the national Natural Science Foundation of China (Grants No. 11675167 and No.11947301) and the national ``Young Thousand People" program.

\end{document}